\documentstyle[12pt]{article}
\begin{document}
\newcommand{\vsone}{\vspace{1cm}}
\newcommand{\be}{\begin{equation}}
\newcommand{\ee}{\end{equation}}
\newcommand{\bea}{\begin{eqnarray}}
\newcommand{\eea}{\end{eqnarray}}
\newcommand{\pr}{\paragraph{}}

\newcommand{\nd}[1]{/\hspace{-0.6em} #1}
\begin{titlepage}
\begin{flushright}
CERN-TH.6514/92\\
ACT-11/92 \\
CTP-TAMU-43/92 \\
\end{flushright}

\begin{centering}
\vspace{.1in}
{\large {\bf The Origin of Space-Time
as $W$ Symmetry Breaking in String Theory }} \\

\vspace{.2in}
{\bf John Ellis}, {\bf N.E. Mavromatos} and {\bf D.V.
Nanopoulos}$^{\dagger}$   \\
\vspace{.05in}
Theory Division, CERN, CH-1211, Geneva 23, Switzerland  \\

\vspace{.05in}
\vspace{.05in}
\vspace{.1in}
{\bf Abstract} \\
\vspace{.03in}
\end{centering}
{\small
\paragraph{}
Physics in the neighbourhood of a
space-time
metric singularity
is described by a world-sheet
topological gauge field theory
which
can be represented as a twisted $N=2$ superconformal Wess-Zumino
model with a $W_{1+\infty} \otimes W_{1+\infty} $ bosonic symmetry.
The measurable $W$-hair associated with the singularity
is associated with Wilson loop integrals around gauge defects.
The breaking of $W_{1+\infty}$ $\otimes $
$W_{1+\infty}$ $\rightarrow $ $W_{1+\infty}$ is associated
with expectation values for open Wilson lines that make the metric
non-singular away from the singularity. This symmetry
breaking is accompanied
by massless discrete  `tachyon'  states that appear as leg poles in
$S$-matrix elements. The triviality of the $S$-matrix in the
high-energy limit of the $c=1$ string model, after renormalisation
by the leg pole factors, is due to the restoration of double
$W$-symmetry at the singularity.}

\par
\vspace{0.4in}
\vspace{0.1in}
\begin{flushleft}
CERN-TH.6514/92 \\
ACT-11/92 \\
CTP-TAMU-43/92 \\
May 1992 \\

\vspace{0.4in}
$^{\dagger}$ {\it Permanent address} :
Center for Theoretical Physics, Dept. of Physics,
Texas A \& M University, College Station, TX 77843-4242, USA,
and  \\
Astroparticle Physics Group,
Houston Advanced Research Center (HARC),
The Woodlands, TX 77381, USA.\\

\end{flushleft}

\end{titlepage}
\newpage
\section{ Introduction and Summary}
\pr
    One of the deepest and most fascinating problems in quantum
gravity is the nature of singularities in space - as inside a black
hole - or time - as at the beginning of conventional Big bang
cosmology. These singularities raise severe challenges for  physics as
we understand it so far: the known laws of physics are no longer
applicable there, and they may cause observable  deviations from the
known laws even in non-singular regions of space-time. For example,
quantum coherence cannot presumably be maintained in any local field
theory in the presence of a space-time singularity \cite{hawk}.
\pr
    String theory provides a powerful framework for addressing and
potentially solving these and other problems in quantum gravity. In
particular, related consistent string models have been formulated
which are able to describe the early stage of Big Bang
cosmology \cite{aben}, and
a spherically-symmetric black hole \cite{witt,emn4}.
It has been shown that quantum
coherence is maintained in the presence of such
a stringy black hole,
thanks to an exact W-symmetry which provides an infinite number of
conserved gauge charges, W-hair, sufficient to label and distinguish
all the black hole states \cite{emn1},\cite{emn2}, \cite{emn3}.
The W-charges of a string black hole are
in principle measurable by scattering experiments similar to
meson-Skyrmion or quark-monopole scattering, or by analogues of
Aharonov-Bohm interference measurements \cite{emnl}.
However, this W-symmetry is
just part of an infinite set of string gauge symmetries \cite{ven,ovr}
whose full
extent and character are not yet understood. The string black hole has
a singularity at the origin, and one could expect a fuller set of
gauge symmetries to be restored there, as occurs at the core of a
gauge monopole \cite{thooft}.
\pr
    In this paper we demonstrate the existence of a double W-symmetry
at the singularity, show how it is broken by Wilson lines down to the
observable W-symmetry of the black hole solution away from the
singularity, demonstrate the relation of tachyon leg-poles to this
symmetry breakdown, and interpret the known triviality of the
high-momentum c=1 string S-matrix as a manifestation of this higher
string symmetry. The singularity at the core of the string black hole
serves as a window on the nature of string before its higher
symmetries are broken by the appearance of an expectation value for
the target-space metric.
\pr
    Our reasoning is as follows. We start from the observation
\cite{witt,eguchi},
that, in string theory, physics in the neighbourhood of a
space-time singularity is described by a topological gauge theory
(TFT) on the world-sheet. It is known that such a TFT can be
represented as a twisted N=2 superconformal
Wess-Zumino model \cite{egyang}. As
such \cite{pope}, it possesses a super-W-symmetry, which includes
$W_{ 1+\infty } \times W_{ 1+\infty }$ as bosonic symmetry, to be
compared with the simple observable $W_{1+\infty}$ symmetry of the
string black hole. The target space-time singularity reflects the
presence of a {\it gauge defect}
on the world-sheet \cite{chinese}, and the observable W-charges
are Wilson line integrals around loops enclosing this world-sheet
defect, which can be viewed as a world-sheet
{\it vortex } . The pure gauge
TFT describes physics at the target-space singularity, where the
physical metric vanishes. A non-vanishing metric corresponds to an
expectation value for an open Wilson line terminating on `matter'
fields of the TFT, analogous to quark-antiquark condensation in
QCD.
Similarly to the way the quark condensate breaks chiral SU(N) $\otimes$
SU(N) $\rightarrow$ SU(N) \cite{chiral},
the `matter' condensate breaks the TFT
$W_{ 1+\infty }  \otimes
W_{ 1+\infty } \rightarrow W_{1+\infty}$. The analogues
of the Goldstone bosons of chiral symmetry breaking are mass-zero
states with discrete momenta, explaining the leg-poles of the string
c=1 model and black hole S-matrices, which were previously
mysterious \cite{gross,pol}  \footnote{Continuum leg
poles were first found in a Wick-rotated version in the
last paper of ref. \cite{aben}.}.
In the high-momentum limit, the c=1 model S-matrix elements sample the
singular region alone, and the higher string symmetry there explains
the triviality found after absorbing the leg pole factors into the
vertex operators for the scattered particles \cite{grosskle}.
\pr
\section{ Symmetries
of the Topological Field Theory at the Singularity }
\pr
    Although the discussion outlined in the Introduction should apply
to any point-like isolated singularity in 4-dimensional target
space-time, so far it can be developed explicitly only for the
singularity at the centre of a spherically-symmetric 4-dimensional
black hole. This is described by the Schwarzschild metric, which can
be written as \cite{wheel}
\be
    ds^2 = -\frac{32M^3 e^{-\frac{r}{2M}}}{r}dudv + r(u,v)^2
(sin^2 {\tilde \theta }d\phi ^2 + d{\tilde \theta }^2 )
\label{sch}
\ee

\noindent where
$u$ and $v$ are Kruskal-Szekeres coordinates. Discarding the radial
part of (\ref{sch})
and making a trivial change of coordinates, one finds that
the metric can be written as a conformally-rescaled form of the
2-dimensional black hole metric \cite{witt}
\be
     ds^2_{2bh}=\frac{1}{1-uv}dudv
\label{sing}
\ee

\noindent which is singular when
$uv = 1$. The 2-dimensional (4-dimensional
spherically-symmetric \cite{emn4})
string black hole is described by an exact
conformal theory which can be written as an $SL(2,R)/U(1)$ coset
Wess-Zumino (WZ) model on the world-sheet \cite{witt}:
\bea
\nonumber
 S_{WZ}  =
 \frac{k}{8\pi} \int d^2 z Tr(\partial _i g \partial _i g^{-1})
+ i \frac{k}{12\pi} \int d^3 x \epsilon ^{ijk} Tr(g^{-1}\partial _i
g g^{-1} \partial _j g g^{-1} \partial _k g ) \\
  +
\frac{k}{2\pi}\int d^2 z Tr (B_{{\bar z}}\sigma _3 g^{-1}
\partial _z g + B_z \sigma _3 {\bar \partial _z } g g^{-1}
+ B_z \sigma _3 B_{{\bar z}} + B_z \sigma _3 g B_{{\bar z}}
\sigma _3 g^{-1} )
\label{wz}
\eea

\noindent where g is a $2 \times 2$ matrix parametrized by
\be
g=\left( \begin{array}{c}
\nonumber          a   \qquad        u   \\
              -v   \qquad         b   \end{array} \right)
\qquad , \qquad ab + uv =1 \\
\label{sl}
\ee

\noindent from
which we see that the space-time singularity $uv = 1$
corresponds
to $ab = 0$. The action (\ref{wz})
is invariant under the U(1) gauge
transformation
\bea
\nonumber \delta g & = &\epsilon (\sigma _3 g + g \sigma _3),
\qquad \qquad \delta B_i = - \partial _i \epsilon  \\
(\delta a & =& 2 \epsilon a, \qquad \delta b =-2\epsilon b ,
\qquad \delta u =\delta v = 0  )
\label{gaug}
\eea

\noindent Eliminating
the gauge field $B_z$ from the action (3), the latter can
be written to leading order in $k$ as \cite{witt}
\be
  S=\frac{-k}{4\pi} \int d^2 x \frac{\partial _i u \partial _i v}
{1-uv}
\label{blhol}
\ee

\noindent where
we see explicitly that the WZ model (\ref{wz}) corresponds to the black
hole metric (\ref{sing})
with its target-space singularity at $uv = 1$.
However,
the WZ model is well-defined even in the neighbourhood of the
singularity \cite{witt,eguchi}.
Parametrizing $u$ and $v$
in the forms $u = e^w,\; v = e^{-w}$,
the action (\ref{wz}) there becomes
\be
S=-\frac{k}{4\pi} \int d^2x D_i a D_i b + i \frac{k}{2\pi}
\int d^2x w \epsilon^{ij}G_{ij}
\label{topol}
\ee

\noindent which
has the form of a U(1) topological gauge field theory (TFT) on
the world-sheet, with matter fields $a,b$, and
$G_{ij}$ is the
gauge field strength.
\pr
    To see the symmetries of this TFT, we recall that it can be
rewritten as a twisted $N=2$ supersymmetric
Wess-Zumino (SWZ) model \cite{egyang}.
This is obtained from the superconformal version of the SL(2,R)/U(1)
model (\ref{wz}):
\be
    S_{susy}= S_{WZ} +
    \frac{i}{2\pi} \int d^2z (Tr \psi D_{{\bar z}} \psi
+ Tr {\bar \psi} D_z {\bar \psi})
\label{susy}
\ee

\noindent where
$\psi$ and ${\bar \psi}$
are coset fermions. The model (\ref{susy}) is twisted by
adding to the stress tensor a piece proportional to the U(1) current,
so that it has zero central charge, one of the supersymmetry operators
is turned into a BRST operator, the fermions are converted into ghost
fields $\alpha$, $\beta$ of spins 0,1, and the action becomes
\be
  S^{twisted}_{WZ}=S_{WZ} + \frac{i}{2\pi}
\int d^2 z Tr(\beta _z D_{{\bar z}} \alpha + {\overline
\beta _{{\bar z}} } D_z {\overline \alpha } )
\label{twist}
\ee

\noindent Physical states are annihilated
by the BRST transformation:
\be
  Q|physical\rangle =0
\label{brst}
\ee

\noindent and
are given by the chiral ring of the $N=2$ theory.
\pr
It will be convenient in the following to use
the twisted $N=2$ $SWZ$ action in the form given by Nojiri \cite{nojiri}:
\bea
\nonumber
S^{(1)} & =& \frac{k}{\pi} \int d^2z tanh^2r (\partial \theta
{\overline \partial } \theta - \partial {\bar \psi } {\bar \psi}
+ \psi {\overline \partial } \psi ) \\
\nonumber   - &2& \frac{sinh r}{cosh^3 r}
(\eta \psi {\bar \partial } \theta
+ {\bar \eta } {\bar \psi } \partial \theta ) +
4 tanh^2 r \eta {\bar \eta }\psi {\bar \psi }   \\
& + & \partial r {\bar \partial } r - \partial {\bar
\eta } {\bar \eta } + \eta {\bar \partial }\eta )
\label{nord}
\eea

\noindent where $r$ and $\theta$ are the physical coordinates
related to the Kruskal-Szekeres coordinates
introduced in equation (\ref{sch}), and $\{ \eta, \psi,
{\bar \eta}, {\bar \psi}\}$ are ghosts.
\pr
Now we come to our first key new point: since this is an SWZ model, it
has a super-$W_{ 1+\infty }$ symmetry.
As such, its bosonic symmetry is
of the form $W_{ 1+\infty }\otimes W_{ 1+\infty }$ \cite{yu,pope},
whose structure can
be seen conveniently via the super-KP hierarchy \cite{yu}:
\be
    \Lambda = \rho ^2 + \sum_{r=0}^{\infty}U_{r} \rho ^{-r-1}
\label{skp}
\ee

\noindent
where $\rho \equiv \partial _{\zeta}
 + \zeta \partial _x $,
$ \rho ^2 = \partial _x $,
 $(\zeta,x)$ are the odd-
and even-parity components of (1,1) superspace,
and $U_i \equiv  v_i(x) + \zeta u_i(x)$. Redefining
\be
   {\tilde u_{2i}}= u_{2i} + v_{2i+1}
\label{redf}
\ee

\noindent
we see that $\{ {\tilde u_{2m}},v_{2r}\} = 0$,
and hence the bosonic currents
$\{ {\tilde u_{2m}}\}$ and $\{  v_{2r} \}$
constitute independent bosonic KP hierarchies.
Hence they together realize a direct product $W_{1+\infty} \otimes
W_{1+\infty}$ bosonic symmetry, whilst the currents $u_{2k+1}$
and $v_{2r}$
 are
fermionic operators which we discuss in the next section.

\section{ W-symmetry Breaking }
\pr
The double $W$-symmetry we have found above is larger
than the single $W$-symmetry of the black hole solution
\cite{emn1,emn2,wu,kir}. Therefore,
our next step is to identify the pattern of symmetry breaking for
this double W-symmetry (cf the breaking of $SU(N)_L \otimes
 SU(N)_R \rightarrow  SU(N)_V$
in QCD), and the mechanism responsible (cf $q{\bar q}$
condensation in QCD). Additional twists are needed, beyond
those leading to (\ref{twist}),
to construct a topological version of
the
$super-W_{1+\infty}$ algebra. This is because, although the currents
$v_{2r}$
are nilpotent, as
follows directly from
their anticommutation relations, the currents
$u_{2r+1}$ become nilpotent only after
twisting as follows \cite{yu}:
\be
 {\tilde u_{2r+1}}= u_{2r+1}- v_{2r+2}
\label{toptw}
\ee

\noindent
This construction leads to two infinite sets of cohomology operators,
and the fermionic currents ${\tilde u_{2r+1}}$ and $v_{2r}$
become ghosts. Thus there
are two BRST operators:
\bea
\nonumber Q_1 &=& \int _y {\tilde u_1 (y) } \\
          Q_2 &=& \int _y  v_0(y)
\label{char}
\eea

\noindent which do not anticommute:
\be
 \{ Q_1 , Q_2 \} = \int _x u_0 (x)
\label{comm}
\ee

\noindent
The physical states are those annihilated by just
one of the two BRST
charges (\ref{char}). It
is precisely the non-anticommutativity property (\ref{comm})
that guarantees that the physical states are not doubled, an
indication of spontaneous symmetry breaking in one of
the sectors.
\pr
We believe that the physical states are those generated
by the $Q_1$
operator, so the physical fields are functionals of
the $v$ sector of the
model. This follows
from the fact
that the unbroken symmetry contains
the entire ${\tilde u}$ sector. This in turn follows
from the observation that
the super-KP operator $\Lambda$, when acting on bosonic functions of x
alone, yields a KP hierarchy
with respect to the coefficients ${\tilde u}$ in
(\ref{redf}). Since
we
are
interested
only in the physical non-ghost sector, we
discard the ghost contributions in (\ref{skp}), and are
left with
\be
\Lambda ^{phys} =\partial _x + \sum _{k'=0} ^{+\infty}
(u_{2k'} + v_{2k' + 1})\partial _x ^{-(k'+1)}
\label{lphys}
\ee

\noindent
It is the coefficients of this operator that appear in higher orders of
the OPE between parafermions of the bosonic WZ coset model
of level parameter $k \ge 2$
\cite{wu,kir},
and generate
the $W_{1+\infty}$ of
the theory that is
measurable in target space and responsible, in our
interpretation, for the maintenance of quantum
coherence \cite{emn1,emn2}.
\pr
   Since the world-sheet, where the $W_{1+\infty} \otimes
    W_{1+\infty} $
symmetry appears in the TFT and is subsequently broken, is
2-dimensional,
the mechanism for
W-symmetry breaking
cannot be simple condensation
accompanied by the appearance of Nambu-Goldstone
bosons \cite{col},
but must be some generalization of Kosterlitz-Thouless vortex
condensation with an infinite set of non-zero order parameters
appearing. The starting-point for a
discussion of this possibility is the configuration
of the U(1) gauge field on the world-sheet in
the model (\ref{topol}).
Consider the following potential for a
vortex-antivortex
configuration placed at the origin and the north pole
of the
world-sheet spherical surface :
\be
\theta =\frac{1}{2} ({\rm Im}  ln z - {\rm Im}  ln {\bar z})
\label{mon}
\ee

\noindent We can define an embedding of the $S^2$
world-sheet in a 2-dimensional
target space $X^{\mu}(z,{\bar z}):\mu = 1,2 $
by the relation \cite{chinese}
\be
2z = e^\omega  - e^{-\omega}
\qquad :
\qquad \omega \equiv  r + i\theta
\label{emb}
\ee

\noindent It is then easy to see that the target-space metric
$ds^2 =
g^{ij}dX_i dX_j = \frac{1}{ 1+z{\bar z}}
dz d{\bar z} $ is
of (Euclidean) black hole type:
\be
     ds^2 = dr^2 + tanh^2 r d\theta ^2
\label{metr}
\ee

\noindent
This demonstrates that a singularity in space-time corresponds to a
topological defect in the U(1) gauge theory on the world-sheet. The
dynamics
of symmetry
breaking and
the values of
the measurable W quantum numbers are associated with
Wilson line integrals in
this gauge theory, as we now see.
\pr
To acquire a physical picture of the breaking
of $W_{1+\infty} \otimes W_{1+\infty} \rightarrow W_{1+\infty}$,
we recall from the parametrization
(4) of the $SL(2,R)$ matrix $g$ that
$ab + uv = 1$, so that a non-zero expectation value for the product
$ab$ corresponds to $uv \ne 1$, and hence a
non-singular region of target-space.
Experience with 2-dimensional gauge theories
tells us that in a TFT such as (\ref{topol}) we should expect
$ \langle ab \rangle \ne 0 $
when $r \ne 0$, which is the case away from
the
world-sheet
defect. Thus gauge dynamics, reminiscent
of that responsible for chiral symmetry breaking in
QCD \cite{chiral}, {\it generates }
space-time away from the singularity.
\pr
To see this formally, it is convenient to use
Nojiri's representation (\ref{nord})
of the $TFT$ action in twisted $N=2$ $SWZ$ form. At the singularity,
$r \rightarrow 0$, we see that it becomes a free field theory,
\be
    S^{(1)} = \frac{k}{\pi} \int d^2 z [ \partial r {\bar \partial r}
- \partial {\bar \eta }{\bar \eta} + \eta {\bar \partial} \eta ]
\label{free}
\ee

\noindent whilst $r \ne 0$ corresponds to a flat direction
of the
the effective potential along which the effective Thirring
four-fermion interaction acquires a non-zero coefficient
(consistently with conformal invariance). The free action (\ref{free})
clearly possesses super-$W_{1+\infty}$ symmetry, which is absent
when the other terms in $S^{(1)}$ (\ref{nord})
are switched on at $r \ne 0$.
\pr
To see how the measurable $W$-charges of the residual $W_{1+\infty}$
are determined we
recall that in higher-dimensional gauge theories formulated
on space-times with non-trivial topologies, gauge symmetry breaking can
be achieved
via non-trivial
boundary conditions for
quantum fields transported around
a Wilson loop enclosing a
defect in the space-time \cite{hosotani}.
Specifically, in theories with a non-Abelian
gauge symmetry G and
\be
    \langle F_{\mu\nu} \rangle =0
\label{vac}
\ee

\noindent
in vacuo in a non-simply-connected space-time, the choice of vacuum
\be
       \langle A_\mu \rangle = V^{\dagger} \partial _\mu V
\label{gauv}
\ee

\noindent
can be physically distinct from the symmetric vacuum
\be
       \langle A_\mu \rangle = 0
\label{zero}
\ee

\noindent
This is achieved by imposing non-trivial boundary conditions for
transport around a closed loop enclosing a defect:
\be
             A_\mu (x+L)   =   U A_\mu (x) U^{\dagger}
\label{twish}
\ee

\noindent
where $U$
is a constant matrix, via a non-zero Wilson loop integral around
the defect:
\be
     W(L) = P exp(i\int _L A_\mu dx^\mu ) U =V^{\dagger} U^{symm} V
\label{wilson}
\ee

\noindent
The gauge transformations $\Omega (x)$ that preserve the vacuum
(\ref{zero}) are
those that satisfy
\be
        [ V(x+L)UV(x)^{\dagger}, \Omega (x) ]=0
\label{co}
\ee

\noindent
so the gauge symmetry is reduced \cite{hosotani}.
\pr
   What about our case? In general in two dimensions there is
tunnelling between the vacua, induced by instanton effects, which
prevents $A_{\mu}$
from taking a
definite local
vacuum expectation
value. The ground-state wave-function is not  a delta-function,
but has smooth support, and no definite vacuum choice is made. The
quantities $U$
and $V$ are not well-defined, and the Wilson integrals (\ref{wilson})
do not act as order parameters for the breaking of the symmetry.
However, in the case of the TFT of interest to us, we believe
on the basis of the absence of instanton effects in the
$c=1$ string model \cite{matrix}, and
because of supersymmetry \cite{affl}, that
tunnelling effects are suppressed, so the above mechanism for symmetry
breaking can be applied to one of the $W_{1+\infty}$ symmetries.
\pr
   Remember that all the discrete quasi-topological states
of the two-dimensional
string in a black hole background
can be regarded as generalized gauge states related to singular
gauge transformations \cite{klepol}. This means that one
can define quantities of the form (\ref{gauv})
for discrete values of the energy
and momentum, which are the only remnants in two dimensions of
string states in higher-dimensional target spaces. The above
symmetry-breaking mechanism applies
away from, but arbitrarily close to,
the defect on the world-sheet, and
breaks down
at the defect,
where the representation (\ref{gauv})
makes no sense. Hence, the symmetry-breaking pattern that
we describe below
does not apply at the target-space singularity, where the
full
symmetry
of the TFT
is recovered. The complete
uncertainty in space-time location of these higher string
states
is to be expected on the basis of the Mermin-Wagner-Coleman
theorem \cite{col}
in two-dimensional space-times, which forbids symmetry
breaking via a local condensate.
\pr
   Our scenario is the following. At the singularity there is no
concept of space-time, only the world-sheet on which the WZ model is
defined. The Wilson loops that
break the symmetry are defined
on the world-sheet as line integrals for the string gauge
states. These loops are well-defined small-radius
($R \rightarrow 0$)
limits of the boundary operators (around closed loops $\gamma$)
introduced in ref. \cite{bound}
\be
      W(L)=lim_{R \rightarrow 0} W(\gamma ) \qquad : \qquad
W(\gamma) =\int _{\gamma} e^{\beta \phi} O(X)
\label{boundaries}
\ee

\noindent where $O(X)$
are matter operators, $\beta$ has the appropriate value to
ensure that
the boundary dimension is one, as required by conformal
invariance,
and the Liouville field
$\phi$ contains a singular logarithmic
piece, $lnz$,
as well as regular parts \cite{bound}.
The logarithmic
part is cancelled, in the small radius limit, by the vanishing
measure of the line integration, so the limit is a well-defined
local operator with discrete energy and
momentum.
The world-sheet defect enables one to impose non-trivial
boundary conditions which are elements of the global
world-sheet
$W_{1+\infty}$
generated
by the $v$
sector of
the super-KP
hierarchy discussed in section 2. This is
consistent with the previously-mentioned
fact that the physical states are constructed out of the
$v$ sector.
The remaining exact W symmetry is realized by the states of the TFT,
which are singular gauge transformations
of the form (\ref{gauv}).
\pr
The
world-sheet $W$-symmetry can be elevated to space-time,
as discussed in \cite{emn4},
where it becomes a local
gauge symmetry \cite{emn1,indians}. Accordingly,
because of the embedding (\ref{emb}) of the
world-sheet in the space-time, the small
loops (\ref{boundaries}) can be regarded
as small space-time integrals, and hence
the above global world-sheet symmetry
breaking mechanism is elevated to the
familiar \cite{hosotani} local gauge symmetry
breaking mechanism.
The
non-trivial boundary conditions are constant group
elements of the full $W_{1+\infty} \otimes
 W_{1+\infty}$, and
 the physical gauge symmetry
 of the space-time away from the singularity
 is due to those generators of this symmetry group that commute
with the Wilson line integrals around the defect on the world-sheet.
This mechanism leaves unbroken the $W_{1+\infty}$
symmetry generated by the ${\tilde u}$ sector of the TFT,
which commutes with the $v$ sector.
\pr
   It should be noted, on the basis of the the commutation relations of
the generators $W_s$
of the $W_{1+\infty}$ algebra constructed out of the
world-sheet currents $v_r$, that there
are always elements with non-vanishing
commutators, i.e., the broken algebra
  is centreless. For instance, exploiting the bi-Hamiltonian structure
of the KP hierarchy in the no-ghost sector \cite{wu2},
we observe that the
anticommutators of the first Hamiltonian structure that
generate $W_{1+\infty}$ contain
\be
 \int dz' \{ W_r (z) ,  W_2 (z') \} = \partial _z W_r (z)
\label{center}
\ee

\noindent
Since the $W_r(z)$ constitute the full set of generators of
$W_{1+\infty}$,
it is clearly always possible to construct boundary condition matrices
(\ref{gauv})
 such that (\ref{co}) is not satisfied, and hence the entire
$W_{1+\infty}$ symmetry group is broken.
\pr
   It is natural to ask what determines the choice of boundary
conditions, which seems arbitrary at first sight. In our picture, they
are determined
dynamically by
the appropriate
choice of vacuum,
though we cannot yet
give a rigorous proof of this statement, since
a complete description of the TFT is not yet available.
However, it should be possible, within the context of string theory, to
understand these boundary conditions as specifying
some sort of sigma-model vacuum configuration. Its determination
 is of course as difficult as any other issue related to lifting the
string vacuum degeneracy, and will not be pursued further in this paper.
We note, however, a close analogy with the problem
of fixing the boundary conditions for the wave-function of the
 Universe at the initial singularity \cite{hartle}.
 Indeed, we would say that this is
the same problem, since the initial cosmological singularity presumably
shares the black hole singularity
features discussed here. The difference
in our approach, compared to that of Hawking,
is that our boundary conditions should be specified on the
world-sheet, with the space-time properties emerging as derived
quantities.

\section{ Leg Poles as Mermin-Wagner-Coleman Bosons }
\pr
   One of the outstanding puzzles of the string c=1 and black hole
models has been the appearance of poles in the external leg momenta of
tachyon scattering amplitudes \cite{gross,pol}.
These are distinct from the massive
discrete states found by factorization of the
tachyon S-matrix, which are known to be
associated with topological states
that are in 1-to-1 correspondence with the generators of the measurable
W-algebra \cite{ms}. The leg poles appear at
light-like momenta, $\sqrt{\alpha '}
p = n^+$,  where $\alpha '$  is the Regge slope, and
$n^+$ is a positive integer, and hence correspond
to discrete massless tachyon states.
The fact that their momenta are
discrete is consistent with the failure of the Goldstone-Higgs mechanism
in two dimensions. In local
field theories, the breaking
of a gauge symmetry does not
yield a Goldstone boson,
since the latter become
the longitudinal components of the vector
bosons, and as such are gauged
out of the physical spectrum. However, at
the special values of momentum
where the leg poles appear one
cannot gauge away longitudinal components of string
gauge states \cite{pol}. Hence
discrete tachyon states appear, which are
non-local in target space, and associated with the breaking of the gauge
symmetry. It
is for this reason that we call them Mermin-Wagner-Coleman bosons.
\pr
The appearance of these Mermin-Wagner-Coleman bosons
can also be seen via the Nojiri's action $S^{(1)}$ (\ref{nord}).
The Thirring interaction term
in $S^{(1)}$ may be expanded with coefficients
in the following series at large $r$:
\be
        \sum _{m=0}^{+\infty} A_m e^{-2mr}
\label{expa}
\ee

\noindent After
appropriate normalisation, we observe that
the terms in this series
correspond (after Fourier transformation)
to background values
of plane-wave tachyons
with definite energies,
at precisely the values that correspond to the different
leg poles \footnote{Because we use the concept of plane waves,
one should be careful in subtracting the background charge
terms from the definition of Liouville energies, when one
defines mass-shell conditions and/or
discusses background values \cite{aben}.}.
\pr
Thirring
interactions are known to preserve conformal invariance
of the pertinent $\sigma$ model
for arbitrary expectation values of the (target space)
scalar fields coupled to them
\cite{yank}. Hence,
from the equivalence of the
vanishing $\beta$-function conditions to
stationary points
of the target space effective action \cite{zam}, we
observe that,
on account of the non-propagating, quasi-topological
nature of the leg-pole states,
each term in the
series (\ref{expa})
corresponds
to an independent {\it flat direction} of the target-space
effective potential
of the $N=2$ superconformal theory.
These are the analogues of Goldstone-Higgs bosons
in higher-dimensions \cite{abk}, but here
they are delocalized massless states. They transform non-trivially
under the broken $W_{1+\infty}$, analogously to pions and
chiral symmetry
in $QCD$.
\pr

\section{Triviality of the High-Momentum Limit of the c=1 String Model
S-Matrix }
\pr
   We are now in a position to interpret the observation that, after
absorbing the external leg poles, the S-matrix of the c=1 string model
is
trivial
in the
high-momentum limit \cite{grosskle}. This is because the
rapidly-varying phase factors of high-momentum tachyons
 do not
 sample the non-singular regions of target space-time, but only
the singular region, in which the $W_{1+\infty} \otimes
 W_{1+\infty}$
symmetry
is so high that no non-trivial
S-matrix elements are allowed.
This is what happens in the topological world-sheet
 models that possess this symmetry, and have no bulk contributions to
the scattering matrix.
\pr
This can
be seen explicitly using the Nojiri's action $S^{(1)}$
(\ref{nord}). In the limit $ r \rightarrow 0$
in which the $SL(2,R)/SO(1,1)$ model becomes the $c=1$ string model,
we retain just the leading term in the expansion (\ref{expa}), and
\be
    S^{(1)}=\frac{k}{\pi} \int d^2z (
    \partial \theta {\bar
\partial } \theta + \partial r {\bar \partial }  r + ... )
\label{lead}
\ee

\noindent which is just the action of a theory of free tachyons.
Hence there is no scattering of the tachyons in the $c=1$ model,
and hence the $S$-matrix $ S \rightarrow 1$ in this limit.
However, we understand from the previous discussion that this
is just an asymptotic statement about the high-momentum
limit in which $\frac{p}{M} \rightarrow 0$ for the mass $M$ of any
quasitopological state. At subasymptotic energies, the $S$-matrix
is non-trivial \cite{sak}, and reflects the double $W$-symmetry,
just as
pion couplings reflect the $SU(N) \otimes SU(N)$ symmetry of
$QCD$.

\section{Comments on the Symmetric Phase of String Theory}
\pr
   We would like to conclude with a few general comments about the
symmetric phase of string theory on the basis of the above discussion.
(1) The
symmetric phase of string theory has often been conjectured to
correspond to a topological field theory \cite{groshigh}. Our analysis
of singularities in string theory confirms
that the relevant $TFT$
is one formulated on the {\it world-sheet}, not in space-time.
(2) The
relevant $TFT$ is a {\it topological gauge field theory},
not a topological gravity theory as one might have thought.
So far at least, the metric on the world-sheet does not
dissolve (but see ref. \cite{green}).
However, defects in the effective gauge theory on the world-sheet
can form, and appear at
space-time singularities.
(3) So far we have only discussed a single, isolated
black hole singularity, corresponding to a single gauge defect
on the world-sheet. This could
be adapted to discuss
the initial cosmological singularity.
(4) One should be able to generalize the
analysis to the case with more defects on the world-sheet,
and to higher genera, corresponding to multi-black-hole solutions.
Such a development would open the way to a more precise
and quantitative description of space-time
foam, in which many microscopic space-time singularities presumably
co-exist. It has not
escaped our notice that the Reissner-Nordstrom
extremity
of
string black holes \cite{witt} suggests compliance with
the `no-net-force' condition, sufficient to ensure
{\it superposition} of the solutions in target space \cite{kan}.
In view of our world-sheet defect interpretation of
target-space singularities, it would be interesting
to exploit the world-sheet version of the `no-net-force'
condition.
(5) In this way one should also be able to discuss
a high-temperature transition to the symmetric phase of string
theory, which is presumably analogous to the unconfined phase
of a two-dimensional gauge field theory.
\pr
\noindent {\Large{\bf Acknowledgements}} \\
\par
The work of D.V.N. is partially supported by DOE grant
DE-FG05-91-ER-40633.
\pr

\end{document}